\documentclass[useAMS,usenatbib]{mn2e}

\usepackage{epsfig}
\usepackage{color}

\title[An MHD study of SN 1006]{An MHD study of SN 1006 and
  determination of the ambient magnetic field direction}
\author[Schneiter et al.]{E. Matias Schneiter$^{1}$\thanks{E-mail:
    mschneiter@gmail.com (EMS); pablo@nucleares.unam.mx (PFV);
    ereynoso@iafe.uba.ar (EMR); fabio@ucolick.org (FDC)}, Pablo F.
  Vel\'azquez$^{1}$, Estela M. Reynoso$^2$\thanks{Member of the
    Carrera del Investigador Cient\'\i fico of CONICET}, and Fabio De
  Colle$^3$\\
  $^{1}$Instituto de Ciencias Nucleares, Universidad Nacional
  Aut\'onoma de M\' exico, M\'exico D.F., M\'exico\\ $^2$ Instituto de
  Astronom\'\i a y F\'\i sica del Espacio, C.C. 67, C.P. 1428, Ciudad
  Universitaria, Buenos Aires, Argentina\\ $^3$ Department of
  Astronomy and Astrophysics, University of California, Santa Cruz, CA
  95064, USA}
\begin{document}

\date{Accepted . Received ; in original form}

\pagerange{\pageref{firstpage}--\pageref{lastpage}} \pubyear{2002}

\maketitle

\label{firstpage}

\begin{abstract}
In this work we employ an MHD numerical code to reproduce the morphology 
observed for SN 1006 in radio synchrotron and thermal X-ray emission. We
introduce a density discontinuity, in the form of a flat cloud parallel
to the Galactic Plane, in order to explain the NW filament observed in 
optical wavelengths and in thermal X-rays. We compare our models with 
observations. We also perform a test that contrasts the radio emitting 
bright limbs of the SNR against the central region, finding additional 
support to our results. Our main conclusion is that the most probable 
direction of the ambient magnetic field is on average perpendicular to 
the Galactic Plane. 

\end{abstract}


\section{Introduction}

SN 1006 is a young supernova remnant (SNR) whose morphology in
radio-continuum can be described as a partial shell with a radius of
$15\arcmin$. Since most of the emission arises in two main bright arcs
towards the NE and SW, this SNR has been classified as bilateral or
barrel-shaped \citep[BSNR][]{kesteven1987,gaensler1998}. Another
characteristic of this source is that the thermal X-ray emission is
uniformly distributed throughout the SNR, and is associated to the
ejected material \citep[see][]{cassam2008,miceli2009}.  Recently,
\citet{acero2007} carried out observations of this object with
XMM-Newton, confirming the presence of a bright filament in the NW
\citep[see][]{dubner2002}, suggesting that the expanding shell is
colliding into a denser medium. Non-thermal X-ray emission is
predominant in the NE and SW almost coincident with the synchrotron
radio emission\citep{cassam2008}.

\citet{kesteven1987} propose two possible mechanisms for the
production of the radio-arcs: an intrinsic one, due for example to the
interaction of internal jets with the SNR shell, and an extrinsic one,
dominated by the characteristics of the surrounding ISM.

Several theoretical works have lately studied the extrinsic mechanism
for the formation of two opposed radio arcs in BSNRs through particle
injection models together with analytical self-similar \citep[for
  recent works see][]{cassam2008,petruk2009} or magnetohydrodynamic
\citep[see][]{orlando2007} numerical simulations.

 \citet{orlando2007} explores, through MHD simulations, different
 scenarios for explaining the asymmetries between the arcs in
 BSNRs. They propose two main mechanisms: the SNR expands in a
 non-uniform medium with an almost uniform magnetic field, and the SNR
 evolves in a uniform ISM with a non-uniform magnetic field. In either
 case, the density gradient or the magnetic field gradient are not
 aligned with the line of sight.  To generate the synthetic
 synchrotron maps, they take into account three possible ways of
 particle injection: isotropic, quasi-parallel, and
 quasi-perpendicular. The quasi-parallel (or perpendicular) nature of
 the injection is determined by the angle between the shock normal and
 the shocked magnetic field ($\phi_{Bs}$). According to
 \citet{orlando2007} only the quasi-perpendicular and isotropic
 injection models reproduce radio images similar to the observed ones
 \citep[in agreement with][]{fulbright1990}. However it is important
 to know the direction of the unperturbed interstellar magnetic field
 (ISMF), in order to decide which injection mechanism is predominant.

\citet{rothenflug2004} carried out an analysis of the radio-continuum
and non-thermal X-ray emission of the remnant of SN 1006. By using an
$R_{\pi/3}$ criterion and based mainly on the non-thermal X-ray
emission distribution, \citet{rothenflug2004} conclude that the ISMF
around SN 1006 is mainly perpendicular to the NE and SW bright radio
arcs (i.e. the ISMF is parallel to the Galactic plane).

A different conclusion was obtained by \citet{petruk2009}, who
  employed the analytical Sedov density and pressure profiles, where
  the aforementioned injection particles and self-similar magnetic
  field configuration are assumed to depend only on the shock
  obliquity through the compression factor in order to model the
  radio-continuum emission from SNRs. They analyse the orientation of
the ISMF based on the azimuthal profile of the radio surface brightness on
both observed and simulated radio maps of SN 1006. Applying
this method, they found that the models that best reproduce both the
observed radio morphology and the azimuthal radio brightness profile
for the chosen aspect angles (angle between the ISMF and the line of
sight) are the isotropic and quasi-perpendicular particle injection
models. Therefore, they conclude that the most likely orientation of
the ambient magnetic field at the galactic latitude of SN 1006 is
parallel to the radio arcs, i.e. perpendicular to the Galactic Plane,
in contrast to the expected direction.

In this work, we follow a similar approach as \citet{petruk2009}, but
employing a 2D axisymmetric MHD numerical simulation and solving for
the magnetic field rather than making a-priori assumptions on its
final configuration. This work has a two-fold objective: to explain
the observed morphology in radio and X-rays, and to infer a possible
structure for the ambient magnetic field, which can decide which
injection particle mechanism is the most important to explain the
non-thermal emission (radio and X-ray) of SNRs.

\begin{figure*}
\centering
\includegraphics[scale=.5]{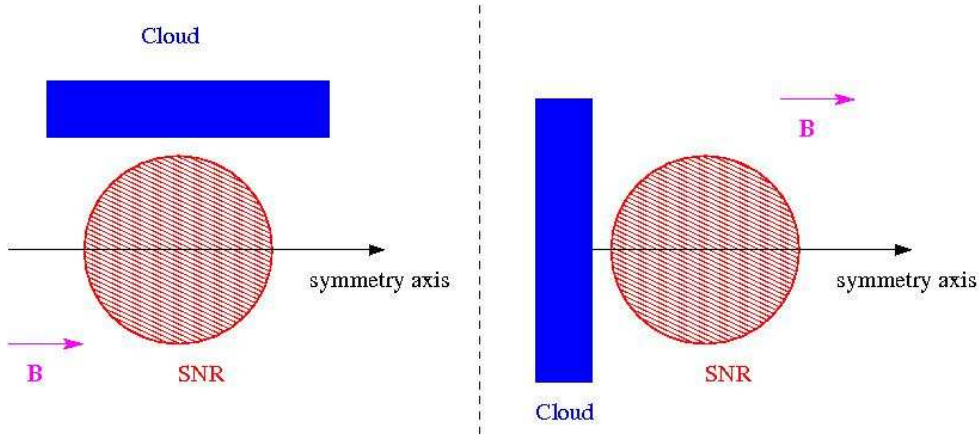}
 \caption{Scheme of the initial numerical setup. The left panel shows
   the setup for model M3, where the magnetic field is parallel to the
   density discontinuity, and the right panel displays the setup for
   M2 where the magnetic field is perpendicular to the
   discontinuity. Notice that due to the axisymmetry assumption, the
   magnetic field has to be parallel to the symmetry axis and, for
   model M3, the density discontinuity is actually a cylinder about
   the SNR.  Model M1 does not include the density discontinuity.}
\label{scheme}
\end{figure*}

\section{Numerical simulation}

\subsection{Code description}

We carry out our simulations using a a two dimensional Eulerian code
(\citealt{decolle2005,decolle2005b,decolle2006}), which solves the
ideal MHD equations, including radiative cooling.

The equations are the following:

\begin{equation}
  \frac{\partial \rho}{\partial t}+\nabla \cdot (\rho {\bf v})=0
\end{equation}
\begin{equation}
  \frac{\partial \rho {\bf v}}{\partial t}+\nabla \cdot (\rho {\bf
    vv}+p_{tot}{\bf I}-{\bf BB})=0
\end{equation}
\begin{equation}
  \frac{\partial e}{\partial t}+\nabla \cdot ((e+p_{tot}){\bf v}-{\bf
    B}({\bf v}\cdot {\bf B}))=-n^2 \Lambda(T)
\end{equation}
\begin{equation}
  \frac{\partial {\bf B}}{\partial t}+\nabla \cdot ( {\bf vB-Bv})=0
\end{equation}

\noindent where ${\bf I}$ is the identity matrix,
$p_{tot}=p_{gas}+B^2/2$ is the total pressure (thermal and magnetic),
$\Lambda (T)$ is the radiative cooling and the remaining symbols have
the usual meaning.  The equations (1-4) represent the mass, moment,
energy and magnetic flux conservation, respectively.

As radiative loss term $\Lambda (T)$ we use a tabulated coronal
cooling function \citep{dalgarno1972}. The radiative cooling function
is switched off at temperatures below $10^4$K.

The constrained transport (CT) method (e.g., \citealt{toth2000}) is
used to conserve $\nabla \cdot B = 0$ to match accuracy.

The MHD Riemann solver uses a second-order Runge-Kutta method for the
time integration and a spatially second-order reconstruction of the
primitive variables at the interfaces (except in shocks where it
reduces at first order).  The fluxes are calculated using the HLL
\nocite{harten1983} method.

\subsection{Initial conditions}

All calculations where carried out on a 2D axisymmetrical grid with a
physical size of $12$~pc and $24$~pc in the $r-$ and $z-$ directions
respectively, with a spatial resolution of $1.5\times 10^{-2}$~pc.  A
uniform interstellar medium of $n_0=5\times 10^{-2} \mathrm{cm}^{-3}$
(Acero, Ballet \& Decourchelle 2007)\nocite{acero2007} with a uniform
magnetic field of $2\mu G$ parallel to the $z$ axis.

The initial conditions consist of a sphere with radius $0.65$~pc,
located at the position $(r_0,z_0)=(0,12)$~pc, containing a total
ejecta mass of $M_\star = 1.4 M_{\odot}$, which is consistent with a
typical type Ia supernova.  An inner region with constant density
containing 4/7 of the ejecta mass was imposed, while the outer region,
containing the rest of the mass, has a density profile following a
power-law density distribution $\rho\propto r^{-7}$
\citep[see][]{jun96}.  This initial density distribution is adequate
for type Ia SNe.

To estimate the initial explosion energy we employed the following
equation \citep{truelove99}:

\begin{equation}
E_{51}=8.87 \times 10^4\! D_{2.2}^2\frac{M^{5/3}_\star}{n_0^{2/3}t_{yr}^2}
\left[ 6.05\!\! \times \!\! 10^{-2}\!\! \left( \frac{R_b^3n_0}{M_\star}
  \right)^{5/6}\!\!\!\!\!\!\!\!\!+0.312 \right]^2
\end{equation}

\noindent where $E_{51}$ is the initial energy in units of
$10^{51}$~erg, $D_{2.2}=D(kpc)/$2.2 kpc is the distance to the object,
$R_b$ is the SNR observed radius in units of pc, and $t_{yr}$ is the
SNR age in years. For the case of SN 1006, $E_{51}=2.05$ considering
$R_b=9.5$~pc and $t_{yr}=1000$. In the simulations, $95\%$ of the
energy was set as kinetic energy and the remaining as thermal energy.

Three models were used for the simulations. In model M1, the remnant
evolves in a uniform ISM with number density $n_0$. In the other
models, and in order to simulate the thermal X-ray emission as well as
the deformation of the NW region of SN 1006, we included an
ISM-density discontinuity at $8.2$~pc from the SNR-centre, with a
density three times higher than the ISM density. Two different
directions for the interstellar magnetic fields were considered by
changing the position of the density discontinuity as shown in figure
\ref{scheme}. In model M2 (M3) the density discontinuity is orthogonal
(parallel) to the symmetry axis (we recall that the symmetry here is
referred to the numerical simulation and not the bright radio and
X-rays arcs of SN 1006). Due to the axisymmetry, the actual density
discontinuity configuration is a ``cylindrical cloud'' around the SNR.

\begin{figure*}
\centering
\includegraphics{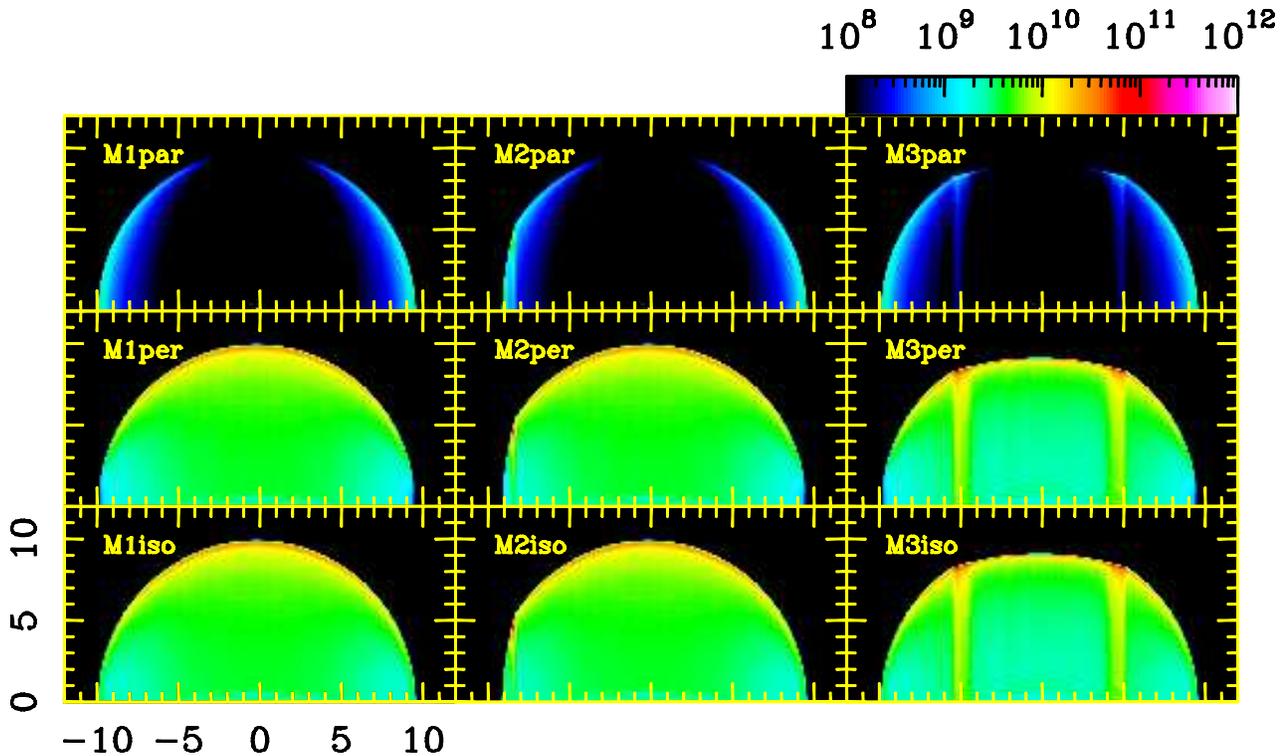}
 \caption{Synthetic synchrotron maps for models M1 to M3 (from left to
   right) and with quasi-parallel (top row), quasi-perpendicular
   (middle row), and isotropic (bottom row) relativistic electron
   density distributions.  The logarithmic colour scale represents the
   synchrotron flux in arbitrary units (same scale is used for all
   models), and the horizontal and vertical length scales are given in
   units of pc. The direction of the ambient magnetic field is always
   along the symmetry axis towards $+x$. The vertical bright lines
   observed in M3 are artifacts introduced by the process for
   obtaining the simulated radio maps from a 2D axisymmetrical
   simulation.}
\label{radio}
\end{figure*}

\subsection{Synchrotron emission simulation}
From the numerical results, we calculated the synthetic radio emission
maps by integrating, along the line of sight, the radio emissivity
given by \citep[see][]{ginzburg1965}

\begin{equation}
i(\nu)\propto K B_{\perp}^{\alpha +1} \nu^{-\alpha}
\label{inu}
\end{equation}

\noindent with $K$ being the normalisation of the electron
distribution, $B_{\perp}$ the magnetic field component perpendicular
to the line of sight, $\alpha$ is the synchrotron spectral index
(which is related with $\zeta$, the index of the electron energy
distribution $N(E)\propto E^{\zeta}$, by $\alpha=(\zeta -1)/2$) whose
value is 0.6 for SN 1006, and $\nu$ is the radio frequency.

The bilateral morphology observed in SNRs, such as SN 1006, can be
generated by the fact that the efficiency of the particle acceleration
at SNR shock front has a systematic dependence on the obliquity angle
$\phi_{Bn}$ between the normal shock and the ISMF.  The literature
bifurcate in two main lines that argue for and against the
quasi-parallel ($\phi_{Bn}=0\degr $) or the quasi-perpendicular
($\phi_{Bn}=90\degr $) models as the most efficient mechanism for
accelerating particles at the remnant shock front. The quasi-parallel
mechanism is associated with the classical diffusive shock
acceleration \citep{blandford87}, while the quasi-perpendicular one
(also called ``shock drift mechanism'') can be a faster process
because of particle acceleration by electric fields along the shock
front \citep[see for example][]{jokipii87}.

\noindent To include all most representative possible cases, in this
  work we study: the isotropic (no $\phi_{Bn}$ dependence), the quasi-parallel
  ($\phi_{Bn}=0\degr $), and the quasi-perpendicular ($\phi_{Bn}=90\degr $) 
particle injection models. Following
\citet{orlando2007} \citep[also][]{fulbright1990}, the isotropic case
was obtained with Eq. \ref{inu} and $K$ being a constant, whereas for
the quasi-parallel and quasi-perpendicular cases, the factor $K$ is
proportional to $cos^2 \phi_{Bs}$ and $sin^2
\phi_{Bs}$\footnote{$\phi_{Bs}$ is the angle between the normal shock
  and the post-shock magnetic field}, respectively. In order to
identify the obtained simulated maps, we have added to M1, M2, and M3,
the words ``iso'', ``par'', and ``per'', for the case of isotropic,
quasi-parallel and quasi-perpendicular particle injection.

\begin{figure*}
\centering
\includegraphics[scale=0.80]{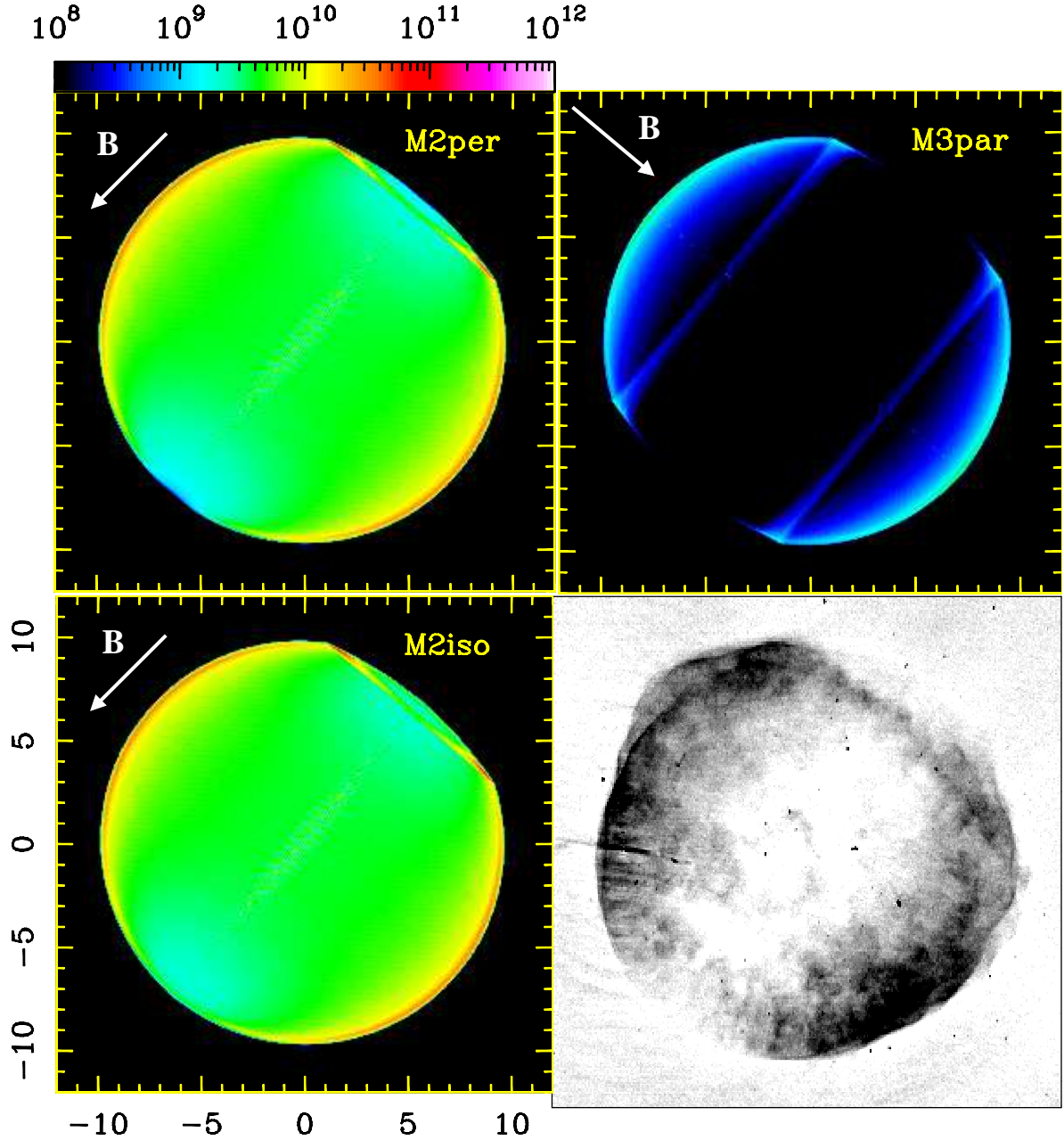}
  \caption{Direct comparison between observation (at 1.5GHz, bottom
    right) and models M3par (top-right), M2per (top-left), and M2iso
    (bottom-left). In order to improve the comparison, simulated
      synchrotron maps from models M2per and M2iso were tilted
      $130^{\circ}$ clockwise in the plane of the sky, while the
      corresponding to model M3par was tilted 40\degr , also clockwise
      in the plane of the sky.}
\label{comparison}
\end{figure*}

\subsection{Thermal X-ray emission}

In order to compare with observations, the X-ray emission coefficient
$j_\nu (n,T)$ was integrated along the line of sight to generate
synthetic emission maps.

The emission coefficient was calculated as $j_\nu (n,T)=n^2 \psi(T)$
(considering the low-density regime), where $n$ and $T$ are the
electronic density and temperature distribution obtained from our
numerical simulations, while $\psi(T)$ is a function that changes
smoothly with $T$. $\psi(T)$ was computed over the energy range
[0.3-5]~keV using the {\sc chianti} atomic database~\footnote{The {\sc
    chianti} database and its associated {\sc idl} procedures are
  freely available at: {http://wwwsolar.nrl.navy.mil/chianti.html},
  {http://www.arcetri.astro.it/science/chianti/chianti.html},
  {http://www.damtp.cam.ac.uk/users/astro/chianti/chianti.html}} and
its associated {\sc idl} software \citep{dere97,landi06}.  The
interstellar absorption was added employing the \citet{morrison83}
model, and taking into account a column density of $N_{H}=7\times
10^{20}\mathrm{cm^{-2}}$ \citep{acero2007,dubner2002}. It was also
assumed that the gas is in coronal ionization equilibrium
\citep[IEQ,][]{mazzotta98}\footnote{The IEQ is an approximation for
  the case of young SNRs such as the SN 1006 giving a lower limit for
  the X-ray flux}.

\section{Results \& Discussion}

Figure \ref{radio} displays a comparison of the synthetic radio
continuum maps among all models, where in M1 the SNR is expanding in a
uniform ISM (left column), in M2 the SNR collides with a density
discontinuity (at a distance of $8.2$~pc from the centre of the SNR)
perpendicular to the symmetry axis of the simulation emulating a
magnetic field perpendicular to the Galactic Plane, and in M3 the
discontinuity is set at the same radial distance ($8.2$~pc) but
parallel to the axis of symmetry, emulating a magnetic field parallel
to the Galactic Plane (the higher density flat cloud, shown in Fig. 1,
is always parallel to the Galactic Plane). The collision between the
SNR blast wave and the density discontinuity occurs at $t\simeq
750$~yr (in the simulation), which is in agreement with previous
estimates \citep[e.g.][]{acero2007}.

By comparing the simulations with observations, we were able to
discriminate some models, keeping M3par, M2per, and M2iso (figure
\ref{comparison}) as possible ones. Models M2par, M3iso, and M3per
were left aside because the flattened shape is parallel to the bright
arcs, which is not observed in the SN 1006 case. The flattened shape
 at the NW and the formation of the two bright arcs are observed in
 M3par, M2per, and M2iso cases. Models M2per and M2iso show a very
 similar brightness distribution as well as intensities, while model
 M3par shows an almost two orders of magnitude lower intensity for the
 maximum at the arcs in comparison to the maximum intensities obtained
 for the quasi-perpendicular and isotropic injection models.
 According to the position of SN 1006 in the Galaxy, one would expect
 a magnetic field perpendicular to the arcs (parallel to the Galactic
 plane), namely, our models M2per and M2iso could be excluded from the
 analysis if this was the case.

\citet{rothenflug2004} employ a $R_{\pi/3}$ criterion which is
basically the ratio between the flux in the interior half of the SNR
and the flux in the limbs. They obtained an $R_{\pi/3}\sim 0.7$ for
the radio-continuum emission of SN 1006, and $\sim 0.3$ and $\sim 0.1$
for the non-thermal X-ray emission of the remnant in the $[0.8-2.]$
keV and $[2.0-4.5]$ keV bands, respectively. \citet{rothenflug2004}
conclude that the bright regions observed at the limbs are probably
polar caps and not an equatorial belt. They also concluded that
acceleration of relativistic electrons proceed preferentially where
the magnetic field is parallel to the shock speed.  We computed the
ratio $R_{\pi/3}$ for synthetic radio-continuum maps for the three
chosen models (see figure \ref{comparison}), and obtained $\simeq
0.7$, $\simeq 0.72$, and $\simeq 0.14$ for M2per, M2iso, and M3par,
respectively. These maps were generated considering an aspect angle of
90\degr\ (i.e. the ambient magnetic field lies on the plane of the
sky). The M3par result exhibit a high contrast in radio emission
between the central region and the polar cap regions, which does not
agree with observations. This high contrast increases up to 3 if an
aspect angle of 11\degr \ (i.e., the symmetry axis is tilted by
79\degr\ with respect to the plane of the sky) is considered when
generating the simulated map. Such aspect angle was calculated by
\citet{petruk2009}.  The values of $R_{\pi/3}$ for models M2per and
M2iso remain unchanged if this criterion is applied to synthetic maps
of these models performed with the aspect angles obtained by
\citet{petruk2009} (64\degr\ and 70\degr\ for M2per and M2iso,
respectively). In other words, the $R_{\pi/3}$ criterion is quite
robust for both models.

Like \citet{petruk2009}, our model is based on the classical MHD
  Rankine-Hugoniot jump conditions for the parallel and perpendicular
  components (with respect to the normal shock) of the ISMF.  This
  could not be valid if the preshock magnetic field was randomized by
  an efficient nonlinear particle diffuse acceleration. In this case
  the unperturbed ISMF can be amplified by orders of magnitude
  \citep{bell2004}. However, \citet{petruk2009} found that, in spite
  of this, their predicted aspect angles remain almost unchanged for
  quasi-perpendicular and quasi-parallel injection models, after
  considering that the compression of a turbulent ISMF does not depend
  on the obliquity \citep[Bohm limit][]{volk2003} and that shocks at
  different initial obliquity under magnetic field turbulent
  amplification become perpendicular inmediatly upstream
  \citep{rakowski2008}.  The case of isotropic injection is discarded
  because it produces synthetic maps with no azimuthal flux
  variations, which is not observed \citep{petruk2009}.

\begin{figure}
\centering
\includegraphics[scale=0.7]{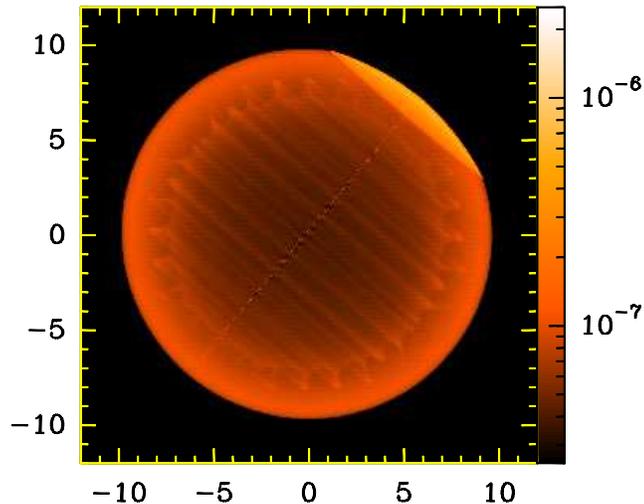}
  \caption{Simulated thermal X-ray emission corresponding to model
    M2. The logarithmic color-bar represents the X-ray flux in units
    of $\mathrm{erg\ s^{-1} \ cm^{-2} \ ster^{-1}}$. The horizontal
    and vertical axes are given in units of pc.}
\label{xray}
\end{figure}

The simulated thermal X-ray map for model M2 is displayed in
fig. \ref{xray}. The symmetry axis ($+\hat{x}$) was clockwise tilted
by 130\degr\ in the plane of the sky, in order to do a direct
comparison with X-ray observations. This image was generated
considering an aspect angle of 90\degr . A bright filament is formed
at the top right region due to the collision with the density
discontinuity. This collision does not affect the morphology of the
radio emission. Rayleigh-Taylor features are observed in the internal
region, close to the contact discontinuity.  Due to the axisymmetry,
the RT fingers become bright lines that cross the SNR
face \footnote{If an aspect angle different of 90\degr\ is employed,
  these bright lines and the bright filament would show an elliptical
  morphology}. In a 3D numerical simulation, the RT features would
give a clumpy appearance to the ejecta emission close to the contact
discontinuity interface \citep[the contact discontinuity density
  distribution was shown in a 3D simulation by][]{miceli2009}.

\begin{figure}
\centering
\includegraphics[scale=0.7]{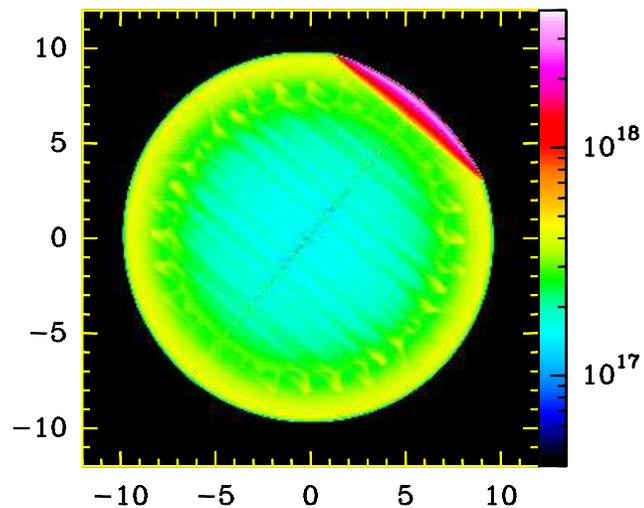}
  \caption{Simulated EM maps corresponding to model M2. The
    logarithmic color-bar represents the EM in units of
    $\mathrm{cm^{-5}}$. The horizontal and vertical axes are given in
    units of pc.}
\label{fem}
\end{figure}

Figure \ref{fem} displays the emission measure ($EM$) distribution
obtained by solving for the integral $\int n_e\ n_H\ dl$, where $dl$
lies along the line of sight. This figure was also clockwise tilted by
130\degr\ in the plane of the sky and the aspect angle is 90\degr . 
The simulated $EM$ can be compared to the parameter $norm$ defined
as $10^{-14}\ EM\ \Omega /4\pi$, where $\Omega$ is the solid angle of
the emitting gas \citep{acero2007}. We obtained an average value for
the bright X-ray filament region of $5\times 10^{-4}
\mathrm{cm^{-5}}$ which doubles the value reported by
\citet{acero2007}.

\section{Conclusions}

In this work we present MHD axisymmetric numerical simulation of the
remnant of SN 1006, taking into account different directions for the
ISMF. Based on the numerical results, synthetic radio and thermal
X-ray emission maps were constructed.

According to our results, and considering the morphology of the radio
and X-rays synchrotron emission, the best fit is given by models M2per
and M2iso, where the magnetic field direction is perpendicular to the
arcs, and therefore perpendicular to the galactic plane. This result
is in agreement with those obtained through analytical models in
\citet{petruk2009}. These results are reinforced when applying the
$R_{\pi/3}$ criterion, since models M2per and M2iso yield a ratio in
accordance to observations (of the order of 0.7) even after correcting
for the aspect angle given by \citet{petruk2009}. At the same time,
the model that could explain the morphology observed with a magnetic
field parallel to the Galactic Plane (M3par) can be rejected with this
test for synchrotron radiation.

Furthermore, our simulated X-ray maps show a bright filament such as
that observed in the NW of SN1006 \citep{acero2007}, and the internal
emission (from the ejected gas close to the contact discontinuity
interface) revealed the RT features. Finally, our estimates of the
$EM$ are in good agreement with the values reported by
\citet{acero2007}, confirming that the collision scenario, while
capable of reproducing the observed thermal X-ray emission, does not
affect the synchrotron emission.

 In summary, our results show that around of the SN 1006 remnant,
   the ISMF direction is mainly perpendicular to the Galactic plane,
   in agreement with previous work \citep{petruk2009}. Based in this
   fact, the best fit with radio observations of this remnant is
   achieved if the injection of the relativistic particles at the
   forward SNR shock wave is isotropic or quasiperpendicular.  This
   result could be different if nonlinear diffuse acceleration process
   takes place at the SNR shock front. In this case the model based on
   isotropic particle injection is no longer valid because the
   predicted synthetic radio maps would show a complete shell
   morphology for any position angle.  Notwithstanding, models based
   on quasiperpendicular particle injection remain valid, with no
   changes in the predicted aspect angle \citep{petruk2009}, if this
   process is independent of obliquity angle $\phi_{Bn}$ (Bohm
   limit).

\section*{Acknowledgments}

Authors acknowledge an anonymous referee for her/his very useful suggestions 
and comments, which help us to improve the previous version of this manuscript.
PFV acknowledges support from grants CONACyT 79744 and DGAPA IN119709.
E.M.R. is partially supported by grants UBACyT X482, PIP
114-200801-00428 (CONICET) and ANPCYT-PICT-2007-00902. Authors thank
Enrique Palacios Boneta (c\'omputo-ICN) for maintaining and supporting the 
linux servers where the numerical simulations were carried out.

\bsp

\label{lastpage}

\end{document}